\documentclass[aps,pra,twocolumn,nofootinbib]{revtex4-2}

\usepackage[colorlinks=true,allcolors=blue]{hyperref}
\usepackage[utf8]{inputenc}
\usepackage{graphicx}
\usepackage{multirow}
\usepackage{amssymb}
\usepackage{amsmath}
\usepackage{physics}
\usepackage{bm}
\usepackage{appendix}
\usepackage{booktabs}
\usepackage{multirow}
\usepackage{bm,color}
\usepackage{subcaption}

\begin{document}

\title{
Searching for scalar field dark matter with hyperfine transitions in alkali atoms
}

\author{V.~V.~Flambaum$^{1}$}
\author{A.~J.~Mansour$^1$}
\author{I.~B.~Samsonov$^{1}$ }
\author{C.~Weitenberg$^{2}$}
\affiliation{$^1$School of Physics, University of New South Wales, Sydney 2052, Australia}
\affiliation{$^{2}$
Institut für Laserphysik, Universität Hamburg, 22761 Hamburg, Germany and \\
The Hamburg Centre for Ultrafast Imaging, 22761 Hamburg, Germany}

\begin{abstract}
Fundamental constants such as masses and coupling constants of elementary particles can have small temporal and spatial variations in the scalar field dark matter model. These variations entail time oscillations of other constants, such as the Bohr and nuclear magnetons, Bohr radius and the hyperfine structure constant. In the presence of an external magnetic field, these oscillations induce hyperfine transitions in atoms and molecules. We determine the probability of magnetic dipole hyperfine transitions, caused by the oscillating fundamental constants, and propose an experiment that could detect the scalar field dark matter through this effect. This experiment may be sensitive to the scalar field and axion dark matter with mass in the range $1\,\mu\text{eV}<m<100\,\mu\text{eV}$.
\end{abstract}

\maketitle

\section{Introduction}

Uncovering the nature of dark matter remains one of the most important unsolved problems in physics. It is hypothesised, in particular, that dark matter may be represented by light bosonic particles, not accounted for in the Standard Model of elementary particles. The leading candidate particles in this class are the axion and axion-like particles, the dilaton-like scalar particle and the dark photon. In this paper, we focus on the scalar particle model with dilaton-like interaction with Standard Model particles that is motivated by superstring theory \cite{Veneziano,DAMOUR1994,Damour1994a,DamourVeneziano,DamourVeneziano2} and chameleon models of gravity, see, e.g., Ref.~\cite{Chameleon} and referenced therein. The mass of the corresponding field and its coupling strengths with the Standard Model fields remain unknown parameters.

If the scalar field describing the cold dark matter is very light, with mass $m_\phi \ll 1$ eV, it may be considered as a classical field oscillating harmonically in every particular point of space, 
\begin{equation}\label{phi}
\phi = \phi_0 \cos( \omega t + \varphi)\,,\quad \omega\approx m_\phi\,,
\end{equation}
where $\varphi $ is a (position dependent) phase.
Assuming that this scalar field saturates all dark matter density, the amplitude $\phi_0$ may be expressed in terms of the local dark matter density $\rho_{\rm DM}\approx 0.4$ GeV/cm$^3$ \cite{DMdensity}, 
\begin{equation}\label{phi0}
    \phi_0 = \sqrt{2\rho_{\rm DM}}/m_\phi\,.
\end{equation}

The interaction of the scalar field $\phi$ with the electromagnetic field $F_{\mu\nu}$ and fermions $f$ (represented by the electron $e$, proton $p$ and neutron $n$ in this paper) is usually considered in the form \cite{Arvanitaki2015,Stadnik2015}
\begin{equation} \label{PhotonInteraction}
    \mathcal{L}_{\text{int}} =  \frac{\phi}{4 \Lambda_{\gamma}} F_{\mu \nu} F^{\mu \nu}
    - \sum_{f = e,p,n} \phi \frac{m_{f}}{\Lambda_{f}} \bar{f} f\,,
\end{equation}
Here $\Lambda_\gamma$ and $\Lambda_f$ are coupling constants of the scalar field to photons and fermions, respectively. In this paper, we propose an experiment which would allow one to determine the values of these coupling constants or find lower limits on their values if the mass of the scalar field appears in the range from 1 to 100 $\mu$eV. 

In Ref.~\cite{oscillation} it was  noted that the interaction (\ref{PhotonInteraction}) of the photon and fermions with the background scalar field (\ref{phi}) implies the variations of the fine structure constant $\alpha$ and masses of the electron $m_{e}$, proton $m_p$ and neutron $m_n$
\begin{align}
        \frac{\delta \alpha}{\alpha} &= \frac{\phi_{0} \cos (m_{\phi}t)}{\Lambda_{\gamma}}\,,
\label{deltaalpha}\\
    \frac{\delta m_{e,p,n}}{m_{e,p,n}} &= \frac{\phi_{0} \cos (m_{\phi}t)}{\Lambda_{e,p,n}}\,. 
\label{deltam} 
\end{align}
Note that, in general, the values of the parameters $\Lambda_\gamma$, $\Lambda_e$, $\Lambda_p$ and $\Lambda_n$ are different. The latter parameter may be related to the corresponding variations of quark masses as in Ref.~\cite{Shuryak2003,FlambaumTedesco}.

The oscillations of the fine structure constant (\ref{deltaalpha}) and masses (\ref{deltam}) imply variations of other physical constants, including the hyperfine structure constant $A$, the Bohr magneton $\mu_B$, and the nuclear magneton $\mu_N$. If the scalar field is ultralight with $m_\phi \ll 1\,\mu\text{eV}$, these variations may manifest themselves as oscillations of frequencies of atomic clocks \cite{Arvanitaki2015,Stadnik2015}, and precision measurements with cold atoms are currently focusing on time-resolved spectroscopy of atomic transitions, see e.g. \cite{Tilburg2015,Hees2016,Kennedy2020,Antypas2021,Kim2022}. However, these variations may drive atomic and molecular transitions if $m_\phi$ appears in a range from $\mu$eV to a few eV.

In this paper, we calculate the probability of hyperfine transitions in atoms due to resonant absorption of the oscillating scalar field (\ref{phi}) and propose an experiment for detection of the scalar field dark matter based on this effect. Such resonant wavy dark matter detectors were previously considered for systems consisting of macroscopic masses \cite{Arvanitaki2016}, molecules \cite{Arvanitaki2018}, atoms inside materials \cite{Zioutas1988,Sikivie2014,Garcon2019,Aybas2021} and recently cold hydrogen atoms for the detection of axion particles \cite{Hyperfine-axion}. As compared with these works, we show that the oscillating scalar field dark matter can cause hyperfine transitions between different $F$ manifolds in atoms. By detecting these transitions with different species of alkali atoms it is possible to explore the range of masses of the scalar field from 1 $\mu$eV to 100 $\mu$eV. This interval is only partly covered by recent constraints from ADMX, which was designed to detect axion dark matter, but also yields constraints for scalar field dark matter \cite{FMST}. The experiment proposed in the present paper would allow for detection of the scalar field dark matter in this frequency interval or impose new limits on the scalar field couplings to the visible matter. Earlier experiments \cite{Antypas2019,Antypas2021,Aharony2021,Savalle2021,Verm,Oswald2022,Tretiak2022,Aiello2022} search for the scalar field dark matter in the lower frequency ranges.

Note that in this paper we focus on the linear in $\phi$ couplings in Eq.~(\ref{PhotonInteraction}). More generally, it is possible to include quadratic in $\phi$  interaction terms with independent coupling constants. In this case  $\phi$ may be either scalar or pseudoscalar (axion) field. Such interactions would imply variations of fundamental constants similar to the ones in Eqs.~(\ref{deltaalpha}) and (\ref{deltam}), but with double frequency. Observational implications of such variations were considered in a series of works \cite{Stadnik2015,Stadnik2015a,Stadnik2016,Stadnik2016a,Hees2018,Stadnik2019,Kim2022}. The proposed in this paper experiment based on hyperfine atomic transitions is suitable for probing the quadratic couplings as well.

The rest of the paper is organized as follows. In the next section, we derive the expression for the oscillating hyperfine magnetic dipole interaction constant via the oscillating fundamental constants. In section \ref{Sec3}, we derive the resonance hyperfine transition rate in alkali atoms induced by oscillating fundamental constants. In section \ref{Sec4} we propose an experiment for detecting scalar field dark matter based on hyperfine transitions in cold atoms and estimate the sensitivity of this experiment to different scalar field couplings. It is noted that this experiment may be sensitive to the axion-electron interaction as well. Section \ref{SecSummary} is devoted to a summary and discussion of the obtained results. In Appendix \ref{AppA}, we collect some details about the coherence time of the scalar field dark matter within the standard halo model.

In this paper we use natural units with $\hbar=c=1$.


\section{Oscillating magnetic dipole hyperfine structure constant}

In this section, we show that oscillations of the fine structure constant (\ref{deltaalpha}) and fermion masses (\ref{deltam}) imply an oscillation of the magnetic dipole moment and the hyperfine coupling constant $A$. 

The nuclear spin contribution to the hyperfine interaction Hamiltonian has the form \cite{Landau1981Quantum}
\begin{equation} \label{Adefinition}
    {H}_{\rm hf} = g_{s} g_{N} \mu_{B} \mu_{N} \frac{8 \pi}{3} |\psi(0)|^2 ({ \bf I} \cdot { \bf J}) \equiv A ({ \bf I} \cdot { \bf J}) \,,
\end{equation}
where $\mu_{N}, g_{N}$ are the nuclear magnetic moment and $g$-factor, $\mu_{B},g_{s}$ are the Bohr magneton and electron $g$-factor, ${ \bf I}$ is the nuclear spin, ${ \bf J}$ is the electron total angular momentum and $\psi(0)$ is the electron  wave function at the nucleus (here we assumed $s$-wave but  further discussion does not depend on this assumption). Equation (\ref{Adefinition}) suggests that $ A \propto g_N\mu_{B} \mu_{N} a_{B}^{-3}$, where $a_{B}$ is the Bohr radius, which dictates $\psi(0)$. However, one has also to take into account relativistic corrections that are significant for heavy nuclei. With the relativistic quantum contributions to the hyperfine structure constant accounted by the factor $K_{\text{rel}}$ calculated in Refs.~\cite{DFW,FlambaumTedesco}, the variation for the hyperfine constant may be cast in the form
\begin{equation}
\frac{\delta A}{A} = \frac{\delta \mu_B}{\mu_B} + \frac{\delta\mu_N}{\mu_N} +\frac{\delta g_N}{g_N}
 - 3\frac{\delta a_B}{a_B} + K_{\text{rel}} \frac{\delta\alpha}{\alpha}\,. 
\label{deltaA1}
\end{equation}
In what follows, we will ignore possible variations of the nuclear $g$-factor because this contribution was shown to be relatively small \cite{Shuryak2003,FlambaumTedesco,Sigma}.

Given the expressions of the fine structure constant $\alpha=\frac{e^2}{ \hbar c}$, Bohr radius $a_B = \frac{\hbar^2}{m_e e^2}$, Bohr and nuclear magnetons, $\mu_B = \frac{e\hbar}{2m_e}$ and $\mu_N = \frac{e\hbar}{2m_p}$, respectively, we have the following variations of these parameters under (\ref{deltaalpha}) and (\ref{deltam}):
\begin{align}
\frac{\delta a_B}{a_B} &= \frac1{\Lambda_a}\phi_0 \cos\omega t\,,&
\frac1{\Lambda_a} &= -\frac1{\Lambda_\gamma} - \frac1{\Lambda_e} \,,
\\
\frac{\delta \mu_{B}}{\mu_{B}} &= \frac1{\Lambda_\mu} \phi_0 \cos\omega t\,, &
\frac1{\Lambda_\mu}&=\frac1{2\Lambda_\gamma} -\frac1{\Lambda_{e}}\,,
\label{oscillatingMagnetons}\\
\frac{\delta \mu_{N}}{\mu_{N}} &= \frac1{\Lambda_N} \phi_0 \cos\omega t\,, &
\frac1{\Lambda_N}&=\frac1{2\Lambda_\gamma} -\frac1{\Lambda_{p}}\,.
\label{oscillatingN}
\end{align}
Substituting these variations into Eq.~(\ref{deltaA1}), we find
\begin{equation}
	\frac{\delta A}{A} = \frac1{\Lambda_A}\phi_0 \cos\omega t\,,
\label{deltaA}
\end{equation}
where 
\begin{equation}
	\frac1{\Lambda_A} = \frac{4+K_{\text{rel}}}{\Lambda_\gamma}
	+\frac2{\Lambda_e} - \frac1{\Lambda_p}\,.
\label{oscillatingA}
\end{equation}

In the next section, we will show that the oscillating magnetons (\ref{oscillatingMagnetons},\ref{oscillatingN}) and the oscillating  hyperfine structure constant (\ref{deltaA}) may drive hyperfine transitions in atoms.


\section{Hyperfine transitions due to variations of fundamental constants}
\label{Sec3}

In this section, we consider atoms with non-vanishing nuclear spin $I$ and one valence electron in the $s_{1/2}$ state such as hydrogen and alkali metals. The total angular momentum is given by the operator ${\bf F} = {\bf I}+{\bf J}$, and the operator of hyperfine interaction is given by Eq.~(\ref{Adefinition}). We assume that the hyperfine constant $A$ in this operator oscillates according to Eq.~(\ref{deltaA}) because of the interaction with the scalar field dark matter. Our goal is to find the probability of hyperfine transitions emerged by the time-dependent hyperfine constant (\ref{deltaA}) and magnetons (\ref{oscillatingMagnetons},\ref{oscillatingN}).

\subsection{No transitions without external magnetic field}

The atomic states may be labeled as $|F,m_F\rangle$ in the basis of commuting operators $\{F^2, F_z, I^2, J^2 \}$. In this basis, the operator of the hyperfine interaction (\ref{Adefinition}) is diagonal, with eigenvalues
\begin{equation}\label{EnergyShift}
    \langle H_{\rm hf} \rangle \equiv  E_F = \frac12A
    [F(F+1) - I(I+1) - J(J+1)]\,.
\end{equation}
When the hyperfine constant oscillates, $A = A(t)$, the energy levels (\ref{EnergyShift}) also oscillate with time, but no transition between these levels occur, as the operator (\ref{Adefinition}) has no off-diagonal elements. Such transitions may be induced by another operator that possesses non-vanishing off-diagonal elements in the basis of hyperfine states.

\subsection{Transitions in magnetic field}

The operator of the interaction of the electron magnetic moment $\boldsymbol{\mu} = -\mu_B({\bf J} + {\bf S})$ with the external magnetic field $\bf B$ is
\begin{equation} \label{HB}
H_B = -\boldsymbol{\mu} \cdot {\bf B} =-\mu_z B\,,
\end{equation} 
where the magnetic field is chosen along the $z$ axis, ${\bf B} = (0,0,B)$. 
The full interaction Hamiltonian is
\begin{equation}\label{Hint}
    H_{\rm int} = H_{\rm hf} + H_{B}\,,
\end{equation}
with the hyperfine interaction Hamiltonian given by Eq.~(\ref{Adefinition}). 

In the basis $|F,m_F\rangle$, the operator of hyperfine interaction (\ref{Adefinition}) is diagonal (\ref{EnergyShift}). The magnetic interaction operator (\ref{HB}), however, possesses both diagonal and off-diagonal matrix elements. In general, these matrix elements may be represented as
\begin{equation}\label{HBmatrix}
\begin{aligned}
    &\langle JIFm_F| H_B |JIF'm_{F'} \rangle = -B(-1)^{F-m_F}
    \\&
    \times\left(
    \begin{array}{ccc}
        F & 1 & F' \\
        -m_F & 0 & m_{F'}
    \end{array}
    \right)
    \langle JIF || \mu_z || JIF' \rangle\,,
\end{aligned}
\end{equation}
where the reduced matrix element of the electron magnetic moment operator is (see, e.g., \cite{Sobelman})
\begin{align}
    \langle JIF || \mu_z || JIF' \rangle &= (-1)^{J+I+F'+1}
    \sqrt{(2F+1)(2F'+1)} \nonumber\\&
    \times\left\{ 
    \begin{array}{ccc}
        J & F & I \\
        F' & J &1
    \end{array}
    \right\}
    \langle J|| \mu_z || J \rangle \,,\\
    \langle J|| \mu_z || J \rangle &= -\mu_B\sqrt{J(J+1)(2J+1)}\nonumber \\&\times
    \left[1+\frac{J(J+1)+3/4 - L(L+1)}{2J(J+1)} \right].
    \label{mu-reduced}
\end{align}\
In what follows, we will focus on atoms in the $s_{1/2}$  ground state. In this case, the reduced matrix element (\ref{mu-reduced}) is 
\begin{equation}
    \langle 1/2|| \mu_z || 1/2 \rangle = -g\sqrt{3/2} \mu_B\,,
\end{equation}
with $g=2$.

Now let us take into account the time-dependent variations of the constants $A$ and $\mu_B$ as in Eqs.~(\ref{oscillatingMagnetons}) and (\ref{deltaA}). This corresponds to the substitutions
\begin{align}
A & \to A\left(1+\frac{\phi_0}{\Lambda_A}\cos\omega t\right)\,,\label{A-oscillate}\\
\mu_B & \to \mu_B \left(1 + \frac{\phi_0}{\Lambda_\mu}\cos\omega t\right)\label{mu-oscillate}
\end{align}
in all matrix elements. Note that the variations of fundamental constants may induce time-dependent variations of the external magnetic field $B$. This effect, however, is similar to the variation of $\mu_B$ as in Eq.~(\ref{mu-oscillate}) and may be fully accommodated by a redefinition of $\Lambda_\mu$.

Equation (\ref{HBmatrix}) shows that the selection rule for transitions induced by the operator (\ref{HB}) with oscillating $\mu_B$ is $\delta m_F =0$, $F'\ne F$. It is convenient to study these transitions in the basis of eigenvectors of the operator (\ref{Hint}),
\begin{equation}
    H_{\rm int}| \psi_i \rangle = E_i | \psi_i \rangle \,.
\end{equation}
Consider, in particular, the lowest energy state $|\psi_-\rangle$ which is mixed with a state $|\psi_+\rangle$ by the operator of the magnetic interaction (\ref{HB}). Explicitly, these states may be written in the basis $|F,m_F\rangle$ as
\begin{widetext}
\begin{equation}
\label{psipm}
    |\psi_\pm \rangle = \frac1{c_{\pm}} 
    \left[
    (4\mu_B B (1-2I) - A (1+2I)^2 \pm (1+2I)d)|F,m_F\rangle +8\mu_B B\sqrt{2I} |F+1,m_F\rangle  
    \right]
\end{equation}
with $F=m_F=I-\frac12$ and
\begin{align}
    d&=\sqrt{16\mu_B^2 B^2 + 8A\mu_B B (2I-1)+A^2(2I+1)^2}\,,\label{d}\\
    c_\pm &= \pm\sqrt{128I\mu_B^2 B^2+[A(1+2I)^2+4\mu_B B(2I-1)\mp (1+2I)d]^2}\,.
\end{align}
\end{widetext}
The corresponding energy eigenvalues are
\begin{equation}
    E_\pm = -\frac14(A\pm d)\,.
\end{equation}

The interaction of the atom with the scalar field is effectively taken into account by the substitutions (\ref{A-oscillate}) and (\ref{mu-oscillate}). Upon these substitutions, the operator (\ref{Hint}) acquires the time-dependent part
\begin{equation}\label{Hprime}
    H_{\rm int}(t) = \phi_0\cos\omega t 
    \left[
        \frac{A}{\Lambda_A} {\bf I}\cdot {\bf J} +
        \frac{2\mu_B B}{\Lambda_\mu} S_z
    \right].
\end{equation}
This operator may be identically rewritten as
\begin{equation}
    H_{\rm int}(t) = \frac1{\Lambda_\mu} H_{\rm int}\phi_0\cos\omega t + \left(\frac1{\Lambda_A} - \frac1{\Lambda_\mu} \right) H_{\rm hf} \phi_0\cos\omega t\,.
\end{equation}
In this form, it is clear that only the last term is responsible for the hyperfine transitions while the first term in the right-hand side is given by a diagonal matrix in the basis $|\psi_i\rangle$ and produces only (time-dependent) energy shifts. Therefore, the rate of the transition from the ground state $|\psi_-\rangle$ to the excited state $|\psi_+\rangle$ on resonance reads
\begin{equation}\label{W30}
    W = \frac{\phi_0^2}{\Gamma} \left(\frac1{\Lambda_A}  - \frac1{\Lambda_\mu} \right)^2 |\langle \psi_+|H_{\rm hf}|\psi_-\rangle|^2\,.
\end{equation}
Here $\Gamma\equiv \delta\omega$ is the width of the frequency distribution of the scalar field dark matter. The corresponding transition probability grows linearly with time because the time-dependent perturbation (\ref{Hprime}) is weak, and the signal coherence time is relatively small.

In general, signal dispersion $\Gamma$ can be written as 
\begin{equation}\label{Gamma}
\Gamma = \gamma m_\phi\,,
\end{equation}
where $\gamma$ is a dimensionless parameter which depends on the dark matter halo model. In this paper, we focus on the standard dark matter halo model with
\begin{equation}
    \gamma \approx 10^{-6}\,,
\end{equation}
see Appendix \ref{AppA} for details. Alternatively, one can consider the caustic ring model \cite{DuffySikivie} which predicts much smaller frequency dispersion, $\gamma \approx 2\times 10^{-10}$. In the latter model, the transition rate (\ref{W30}) is significantly enhanced, leading to stronger constraints on $\Lambda$'s but requiring tighter constraints on the control of the magnetic field.

Calculating the matrix element of the hyperfine operator using the explicit form of the wave functions (\ref{psipm}) we find the transition rate (\ref{W30}) in the form
\begin{equation}
\label{Wtemp}
    W= \frac{\phi_0^2}{\Gamma} \left(\frac1{\Lambda_A}  - \frac1{\Lambda_\mu} \right)^2
    \frac{8IA^2 \mu_B^2 B^2}{ d^2}\,.
\end{equation}
Note that the resonance angular frequency of this transition is
\begin{equation}\label{TransitionFrequency}
   m_\phi = \omega = E_+ - E_- = \frac12 d\,,
\end{equation}
with $d$ given by Eq.~(\ref{d}). Making use of Eqs.~(\ref{phi0}) and (\ref{Gamma}), the transition rate (\ref{Wtemp}) may also be cast in the form
\begin{equation}
    \label{W}
    W = \frac{4 I A^2\mu_B^2B^2 \rho_{\rm DM}}{\gamma m_\phi^5} \frac1{\Lambda_{A\mu}^2}\,,
\end{equation}
with
\begin{equation}\label{LambdaAmu}
    \frac1{\Lambda_{A\mu}} \equiv \frac1{\Lambda_A} - \frac1{\Lambda_\mu}=\frac{3.5+K_{\rm rel}}{\Lambda_\gamma} + \frac{3}{\Lambda_e} - \frac1{\Lambda_p}\,.
\end{equation}
Here we made use of the relations (\ref{oscillatingMagnetons}) and (\ref{oscillatingA}). Thus, the experimental measurements of the hyperfine transitions in atoms allow one to explore this particular combination of coupling constants.

Note that numerical values of the constant $K_{\rm rel}$ were calculated in Refs.~\cite{DFW,FlambaumTedesco} for a number of atoms. In particular, it was found that $K_{\rm rel} = 0.34$ for Rb, $K_{\rm rel} = 0.83$ for Cs, and it is close to zero for lighter alkali atoms.

\subsection{Numerical estimates}

\begin{table}
\begin{center}
\begin{tabular}{c|c|c|c|c|c|c}
\multirow{2}{*}{ }&\multirow{2}{*}{$I$}&\multirow{2}{*}{$A$, $\mu$eV}& \multicolumn{2}{|c|}{$B=0.005$\,T} & \multicolumn{2}{|c}{$B=0.1$\,T}
\\\cline{4-7}
 &  & & $\omega$, $\mu$eV & $W_0$, s$^{-1}$ &  $\omega$, $\mu$eV & $W_0$, s$^{-1}$ \\\hline\hline
H & 1/2 & 5.87 & 5.90 & 3.77 & 13.0 & 29.3\\
D & 1 & 0.903 & 1.64  & 107 & 12.1 & 1.97 \\
$^{6}$Li & 1 & 0.629 & 1.26 & 194 & 11.9 & 1.03 \\
$^{7}$Li & 3/2 & 1.66 & 3.65 & 10.0 & 13.5 & 5.68 \\
$^{23}$Na & 3/2 & 3.66 & 7.63 & 1.22 & 16.5 & 10.3 \\
$^{39}$K & 3/2 & 0.955 & 2.26 & 36.7 & 12.6 & 2.65\\
$^{41}$K & 3/2 & 0.525 & 1.43 & 108 & 12.1 & 0.98 \\
$^{85}$Rb & 5/2 & 4.18 & 12.9 & 0.19 & 22.0 & 5.28 \\
$^{87}$Rb & 3/2 & 14.1 & 28.6  & 0.02 & 35.5 & 3.33\\
$^{133}$Cs & 7/2 & 9.50 & 38.4 & 0.006 & 47.3& 0.83\\\hline
\end{tabular}
\end{center}
\caption{Estimates of the hyperfine transition rates $W_0$ due to oscillating fundamental constants in alkali atoms calculated according to Eq.~(\ref{W0}). The values of the hyperfine constants $A$ in alkali atoms are taken from Refs.~\cite{Arimondo1977,Tt,Allegrini2022}. The transition frequency $\omega$ is estimated through Eqs.~(\ref{TransitionFrequency}) and (\ref{d}).
}
\label{Tab}
\end{table}

For numerical estimates, it is convenient to represent Eq.~(\ref{W}) in the form 
\begin{equation}
    W = W_0  \left(\frac{\rho_{\rm DM}}{0.4\,{\rm GeV/cm}^3}  \right)
    \left(\frac{10^{-6}}{\gamma} \right)
    \left(\frac{1\,\rm GeV}{\Lambda_{A\mu}} \right)^2\,,
\end{equation}
where
\begin{equation}
\label{W0}
    W_0 = \frac{1.6\times 10^6 IA^2\mu_B^2B^2}{1\,{\text{GeV}}\, {\rm cm}^3\, m_\phi^5}\,.
\end{equation}
Numerical estimates of this quantity are given in Table~\ref{Tab} for alkali atoms. We present the results both for weak ($B=0.005$\,T) and relatively strong ($B=0.1$\,T) magnetic fields. Varying the magnetic field between these values it is possible to cover the region of frequencies from $\omega = 1.3\,\mu$eV to 100 $\mu$eV by considering different atoms. In this range, the smaller frequencies may be probed with light atoms such as Li whereas to study the high frequency region heavier atoms such as Rb and Cs should give a better sensitivity. From a practical side, choosing the atomic species lithium and rubidium with their two isotopes $^6$Li, $^7$Li and $^{85}$Rb, $^{87}$Rb already covers the considered mass range with a reasonable sensitivity.

\section{Estimate of experimental sensitivity}
\label{Sec4}

\subsection{Design of the experiment}

In the following we estimate the sensitivity of the proposed technique. An important point is the detection of single excited atom at near zero background, reminiscent of gravitational wave detectors working on the dark fringe of the interferometer, i.e., combining very large intensities with single photon detection in the light detector.

Laser cooled atoms allow for a controlled preparation of spin-polarized samples in the lower $F$ manifold via optical pumping as well as a sensitive selective detection of single atoms in the upper $F$ manifold. The search for a hyperfine-changing signal is moreover not sensitive to imperfect control of the Zeeman sublevels, which might arise from imperfect optical pumping or from spin-changing collisions between atoms. Starting from the lower hyperfine level also excludes dipolar relaxation processes, and the natural lifetime of the upper hyperfine level is much longer than all relevant time scales.

\begin{figure*}[!]
	\setlength{\tabcolsep}{5pt}
	\begin{tabular}{p{4cm}p{4cm}p{4cm}p{4cm}}
	\includegraphics[width=3.8cm]{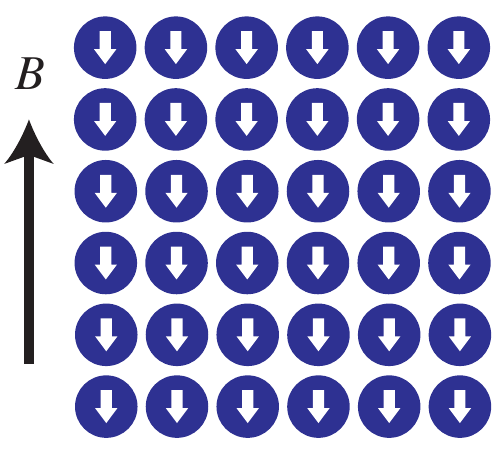}
	& 	\includegraphics[width=3.8cm]{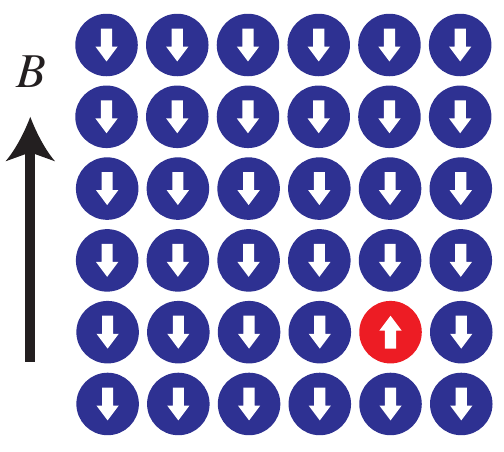} & 	\includegraphics[width=3.8cm]{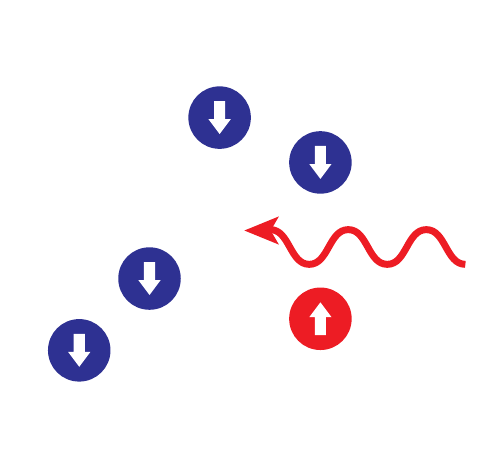} &
		\includegraphics[width=3.8cm]{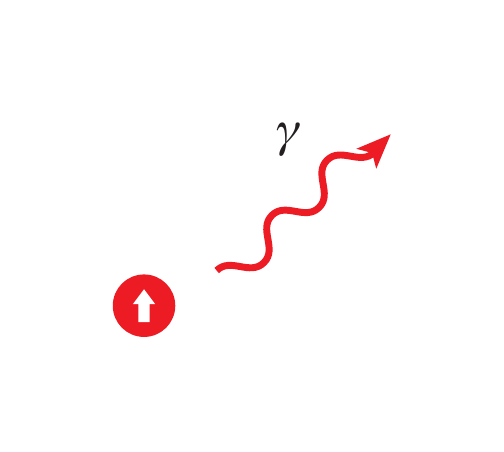} \\
		$N_{\rm at}$ of atoms are prepared in the lower $F$ manifold of hyperfine levels and held in an optical lattice for the probe time $t_{\rm integ}$. & 
		Scalar field dark matter interacting with a static magnetic field $B$ induces an atomic transition to an upper $F$ manifold of hyperfine levels. &
		Push out of atoms in the initial state via state-selective resonant light. &
The remaining atom is detected by inducing fluorescence on an optical transition producing many photons, which are detected on a camera.
	\end{tabular}
	\caption{Sketch of the experimental protocol of the resonant dark matter detector based on cold atoms. 
	}
    \label{fig:scheme}
\end{figure*}

Large alkali atom numbers in the range $N_{\rm at}\sim10^{10}$ can be prepared in a magneto-optical trap (MOT). In the regime of few atoms, the same MOT setup also allows for single atom detection via fluorescence count and careful elimination of stray light \cite{Hu1994,Alt2003,Serwane2011,Hume2013}. The selective detection of the hyperfine-excited atoms can be achieved by selectively removing the atoms in the initial $F$ state via a resonant push out pulse before recapturing the remaining atoms in the MOT or imaging them in an optical lattice \cite{Bakr2010,Sherson2010}. Imperfect push out will lead to a small background signal. Single-atom resolved detection produces clear fluorescence steps per atom, which are larger than the photon shot noise or drifts of the fluorescence light intensity. This procedure should allow one to detect single hyperfine excitations in a large ensemble. Hyperfine-selective resonant push out should be advantageous compared to spatial Stern-Gerlach separation, because it can be done faster and without applying magnetic field gradients. A sketch of the corresponding experimental procedure is given in Fig.~\ref{fig:scheme}.

To let the atoms evolve solely under the influence of the dark matter coupling, one needs to switch off the MOT for an integration time $t_{\rm integ}$ and switch the anti-Helmholtz configuration of the magnetic field coils with a Helmholtz configuration for a maximally homogeneous magnetic field in the interaction volume. Scanning the mass range requires feasible magnetic field strengths up to 2000 G. At this point one can add an optical lattice, which allows for longer integration time by holding the atoms against gravity and prevents collisions between them. Deep optical lattices of sufficient sizes could be realized in bow-tie configuration lattice \cite{Sebby-Strabley2006} or with the help of enhancement resonators \cite{Park2022}. Practical integration times will be in the range of $t_{\rm integ}=10\,$s, to keep it short against atom loss from background gas collisions, but long compared to typical few-second MOT loading times. The integration times are much longer than the coherence time of the expected dark-matter drive, which is allowed for the rate model of incoherent drives considered here. 

It is important to suppress all noise sources in order to allow for a clear identification of the dark-matter signal. Using far off-resonant optical traps allows one to decouple the signal on internal state transitions from the noise on external degrees of freedom, such as thermal motion of the atoms or intensity noise on the optical traps. Furthermore, the trapping frequencies can be kept off-resonant from the hyperfine transitions. The targeted sensitivity corresponds to a relative modulation of the magnetic moment of $10^{-15}$, and false signals from a similar modulation of the magnetic field strength will avoided. Taking into account the sub-Hertz linewidth of the hyperfine transition, relative noise levels of the current in the magnetic field coils of $<10^{-15}/\sqrt{\rm Hz}$ need to be reached, which seems realistic in the relevant frequency range of MHz to GHz. In addition, a $\mu$-metal shielding of the experiment region can suppress magnetic field noise to the low rms $\mu$G level \cite{Farolfi2019}. The value of the magnetic field should furthermore be actively stabilized to $10^{-6}$ during the integration time.

\subsection{Sensitivity to $\Lambda_{A\mu}$}

\begin{figure*}[tbh]
\begin{tabular}{cc}
	\includegraphics[width=8.5cm]{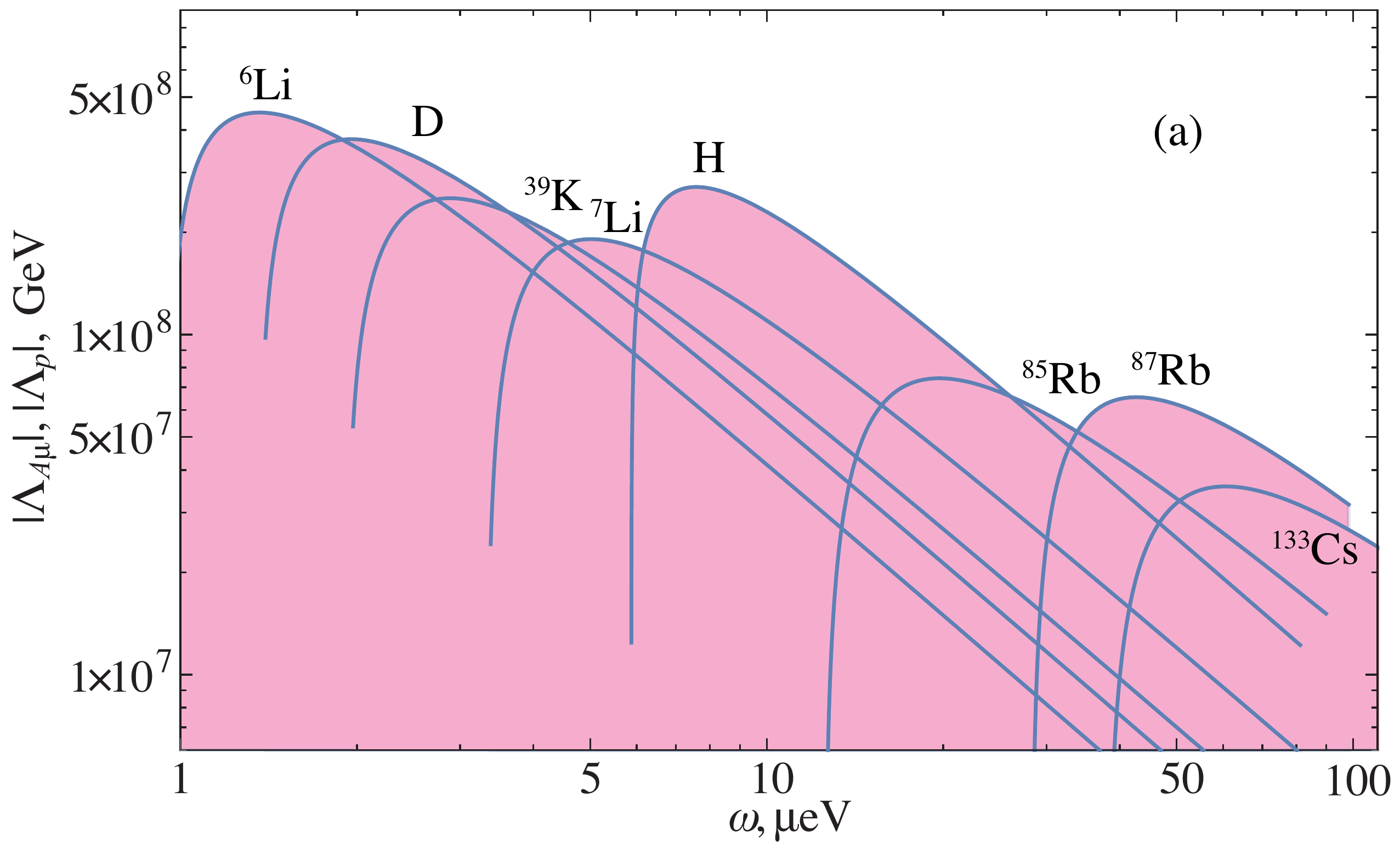} & 
	\includegraphics[width=8.5cm]{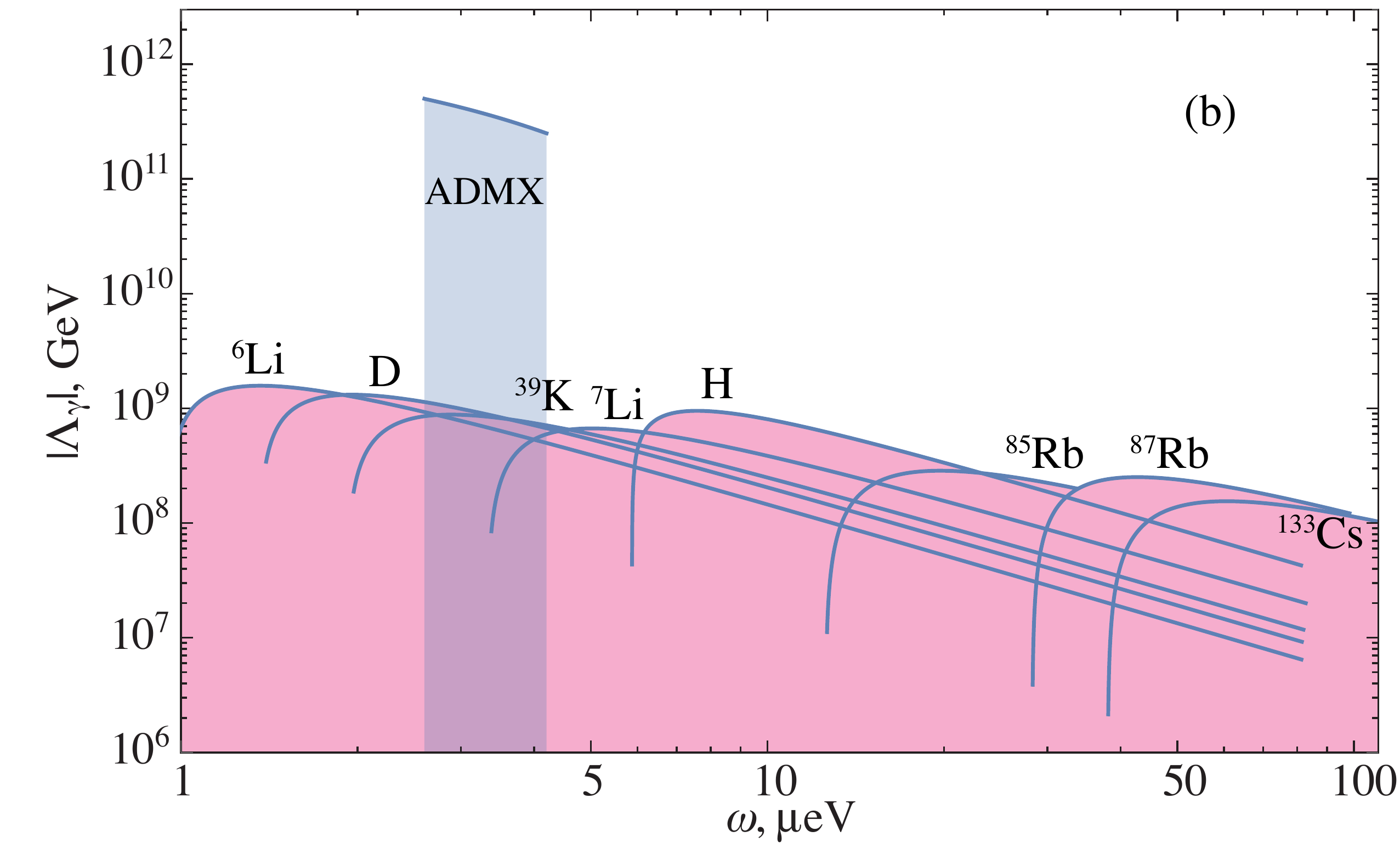} \\
		\includegraphics[width=8.5cm]{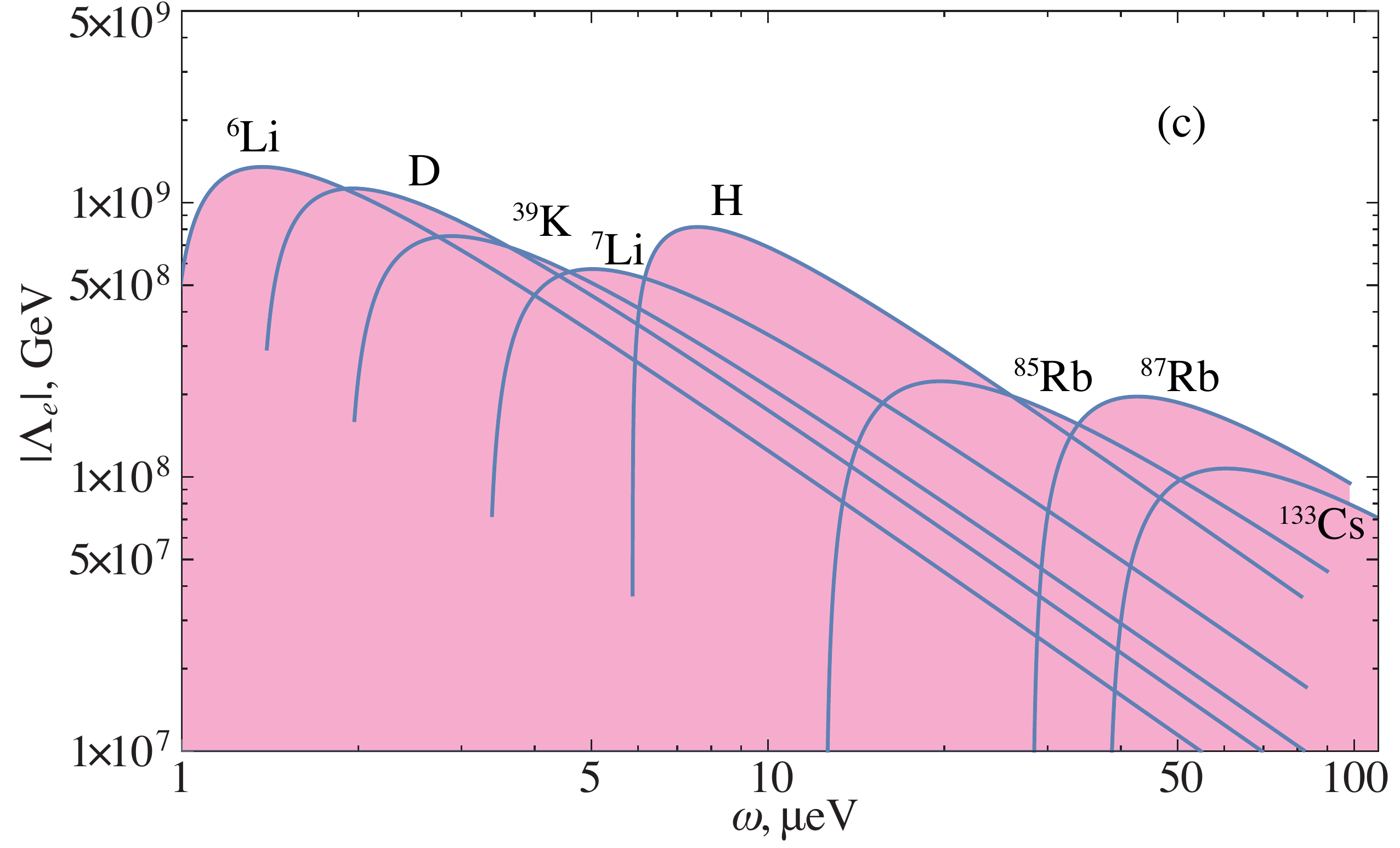} 
		& \includegraphics[width=8.5cm]{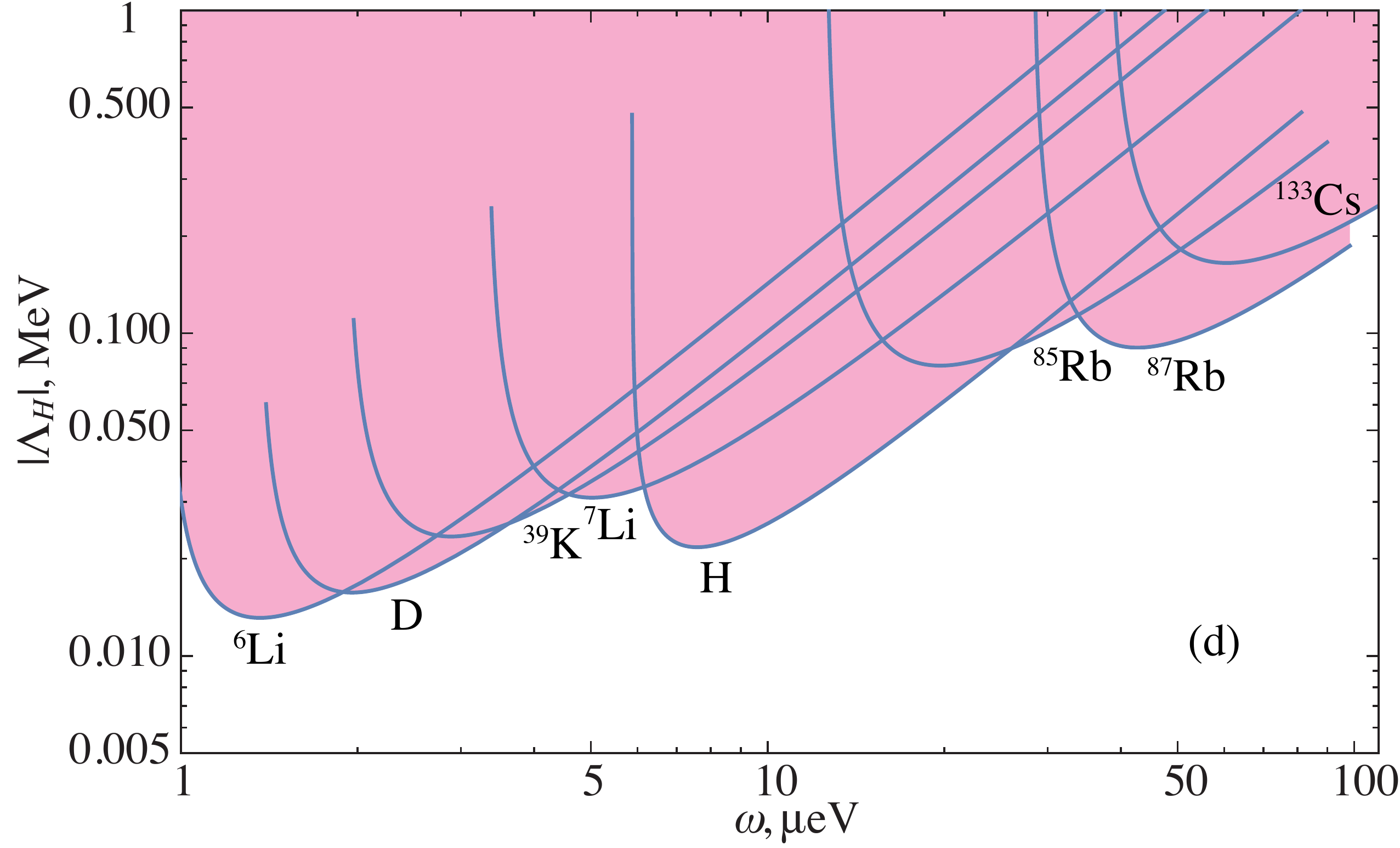}
\end{tabular}
	\caption{(a) Projected sensitivity of the atomic hyperfine transition experiment to the combined coupling constant $\Lambda_{A\mu}$ estimated with the use of Eq.~(\ref{P}) assuming the integration time $t_{\rm integ}=10$ s, number of atoms $N_{\rm at}=10^{10}$ and number of repetitions at each frequency bin $N_{\rm rep}=10^4$. Plots (a)-(d) represent also the sensitivity to the coupling constants $\Lambda_p$, $\Lambda_\gamma$, $\Lambda_e$, and $\Lambda_H$, respectively, with excluded regions colored in pink. In plot (b), we present also the existing limits from ADMX experiment found in Ref.~\cite{FMST}. Note that $1\,\mu\text{eV}=2\pi \times 242\,\text{MHz}$.}
    \label{FigSensitivity}
\end{figure*}

Following the above considerations we consider only the atomic shot noise to estimate the sensitivity of the technique, i.e. we find the parameters at which the probability $P$ to detect an excited atom becomes of order unity: 
\begin{equation}\label{P}
P = W_0 t_{\rm integ} N_{\rm at} N_{\rm rep}\left(\frac{1\,\rm GeV}{\Lambda_{A\mu}} \right)^2,
\end{equation}
with $W_0$ given by Eq.~(\ref{W0}). By setting $P=O(1)$, this equation allows us to estimate the sensitivity of an experiment with $N_{\rm at}=10^{10}$ atoms performing $N_{\rm rep}=10^4$ cycles of measurements with the integration time $t_{\rm integ}=10$ s within each cycle. This corresponding sensitivity is shown in Fig.~\ref{FigSensitivity}. The total measurement time to reach this sensitivity is about 28 hours at each frequency bin.

The relative line width of the signal is $10^{-6}$ (Appendix A.3), so efficiently scanning a large frequency range requires $10^6$ measurements. Long measurement campaigns over one year are possible, because around-the-clock measurements with cold atoms are routinely performed. For scanning a large range within one year, one needs to reduce the number of measurements per frequency bin $N_{\rm rep}$ by a factor of $10^3$ compared to the estimate above and work with a lower sensitivity reduced by a factor $\sqrt{10^3}$.

In our sensitivity estimates we assume the same values of the parameters $N_{\rm at}$ and $N_{\rm rep}$ for all atoms, although there may be technical limitations on these parameters for some atomic species. In particular, for H and D it is challenging to achieve the assumed above number of cold atoms in a trap.

\subsection{Sensitivity to other coupling constants}

The proposed experiment is sensitive to a very specific combination of photon $\Lambda_\gamma$, lepton $\Lambda_e$ and proton $\Lambda_p$ couplings (\ref{LambdaAmu}). Either of these coupling constants may give the leading contribution to the transition rate, and the proposed experiment will not allow one to differentiate these contributions. Assuming that only one of these couplings is significantly stronger than the others, we estimate the experimental sensitivity to this particular coupling. The corresponding plots are presented in Fig.~\ref{FigSensitivity} (a), (b) and (c), for $\Lambda_p$, $\Lambda_\gamma$ and $\Lambda_e$, respectively. In these plots, the curves of the individual atomic species start at the zero-field hyperfine transition and assume Zeeman levels, for which the resonance frequency is shifted to larger values for increasing magnetic field. As expected, the sensitivity first grows with the magnetic field strength and then reaches a maximum at magnetic field strengths intermediate between the Zeeman and the Paschen-Back regime.

For the photon coupling $\Lambda_\gamma$ in Fig.~\ref{FigSensitivity} (b), we present also the constraints from the ADMX experiment found in Ref.~\cite{FMST}. Although these constraints are more than by two orders stronger than the sensitivity of the proposed in this paper experiment, the ADMX experiment is focused on a relatively narrow region from 2.6 to 4.2 $\mu$eV.

It should also be noted that the sensitivity of the proposed experiment is significantly lower than the existing constraints from the experiments probing the violation of the equivalence principle \cite{EPtest}, see Refs.~\cite{Hees2018,Leefer2016}. However, the proposed experiment would allow for an independent detection of the scalar field dark matter with the use of hyperfine transitions in atoms.

\subsection{Sensitivity to the Higgs field coupling}

The scalar field $\phi$ may have linear coupling to the Higgs doublet $H$,
\begin{equation}\label{HiggsCoupling}
    {\cal L}_{\rm int}^H = -\Lambda_H \phi H^\dag H\,,
\end{equation}
where $\Lambda_H$ is a dimensionful coupling constant. In Refs.~\cite{Stadnik2016,Pospelov2010} it was shown that this coupling of the Higgs field with the background scalar field (\ref{phi}) implies the following oscillation of the fundamental constants:
\begin{equation}
\label{HiggsVariations}
\begin{aligned}
    \frac{\delta\alpha}{\alpha} &= \frac{\Lambda_H \alpha}{2\pi m_H^2}\phi_0 \cos(m_\phi t)\,,\\
    \frac{\delta m_e}{m_e} &= -\frac{\Lambda_H}{m_H^2}\phi_0 \cos(m_\phi t)\,,\\
    \frac{\delta m_p}{m_p} &= -\frac{\Lambda_H b}{m_H^2}\phi_0 \cos(m_\phi t)\,,
\end{aligned}
\end{equation}
where $m_H=125$ GeV is the mass of the Higgs field and $b\approx0.2-0.5$ \cite{b}. For our estimates we will assume the central value of the latter parameter, $b=0.35$.

Assuming that the Higgs coupling (\ref{HiggsCoupling}) is the only source of variations of fundamental constants, we find that the variations (\ref{HiggsVariations}) induce atomic hyperfine transitions with the rate (\ref{W}), where the effective coupling $\Lambda_{A\mu}$ is expressed in terms of $\Lambda_H$ as
\begin{equation}
    \frac1{\Lambda_{A\mu}} = \frac{\Lambda_H}{m_H^2}
    \left[ (3.5+K_{\rm rel})\frac{\alpha}{2\pi} -3 +b \right].
\end{equation}
Substituting this expression into Eq.~(\ref{P}), we find the sensitivity of the proposed experiment to the Higgs coupling $\Lambda_H$, see Fig.~\ref{FigSensitivity} (d). To the best of our knowledge, the coupling constant $\Lambda_H$ has not been constrained for the scalar field with the masses in the interval $1\,\mu\text{eV}<m_\phi<100\,\mu\text{eV}$ yet.


\subsection{Sensitivity to axion dark matter}

The same experiment based on detection of hyperfine transitions in alkali atoms may be sensitive to the axion dark matter. The coupling of the axion $a$ to a Dirac fermion $f$ has the general form
\begin{equation}
    {\cal L}_{a\bar f f} = -\frac{g_{aff}}{2m_f} \partial_\mu a \bar f \gamma^\mu \gamma^5 f\,,
    \label{Laff}
\end{equation}
with $g_{aff}$ the dimensionless coupling constant and $m_f$ the fermion mass. For the QCD action, this coupling is related with axion decay constant $f_a$ as $g_{aff}/m_f = g_f/f_a$, where $g_f$ is a model dependent coefficient of order of 1. In this paper, however, we consider general axion-like particles with independent coupling $g_{aff}$.

For non-relativistic electrons, the interaction (\ref{Laff}) implies the following Hamiltonian \cite{GrahamRajendran,SF2014,Sikivie2014}
\begin{equation}
    H_{aee} = \frac{g_{aee}}{m_e} \left(\vec S\cdot \nabla a +\frac{\vec p\cdot \vec S}{m_e}\partial_t a\right)\,,
    \label{Haee}
\end{equation}
where $\vec p$ and $\vec S= \frac12\vec\sigma$ are the operators of momentum and spin of the electron, respectively. In atoms, the first term in Eq.~(\ref{Haee}) is responsible for magnetic M1 transitions while the last one causes parity-changing transitions with $\Delta j=0$ and $\Delta l =1$. Focusing on the experiment based on the M1 hyperfine transitions in atoms, we  will ignore the last term in Eq.~(\ref{Haee}).

Similar to the scalar field $\phi$ in Eq.~(\ref{phi}), the axion dark matter may be described by a classical oscillating field of the form
\begin{equation}
    a = a_0 \cos(\omega t-\vec k\cdot \vec x)\,,
\end{equation}
with energy $\omega$ and momentum $\vec k= m_a \vec v \approx \omega \vec v$. Here $m_a$ is the axion mass and $v\sim 10^{-3}c$ is the galactic halo axion velocity in the laboratory frame. With this field, the interaction Hamiltonian (\ref{Haee}) may be written as
\begin{align}
    H_{aee} &= V \sin(\omega t-\vec k\cdot \vec x)\,,\\
    V &= g_{aee}a_0 (m_a/m_e) \vec S \cdot \vec v\,.\label{V}
\end{align}
On resonance, this interaction can cause an atomic transition with the transition rate
\begin{equation}
\label{W46}
    W= \frac{|\langle f | V|i\rangle|^2}{\Gamma}\,,
\end{equation}
where $\Gamma$ is the axion energy dispersion similar to the one in Eq.~(\ref{Gamma}). 

In the experiment proposed above for detection of the scalar field dark matter, the initial and the final states are given by $|i\rangle = |\psi_-\rangle$ and $|f\rangle = |\psi_+\rangle$, respectively, with $|\psi_\pm\rangle$ given in Eq.~(\ref{psipm}). For these states, we find the matrix element of the operator (\ref{V}), 
\begin{equation}
\label{Velement}
    |\langle \psi_+ | V |\psi_-\rangle|^2 = \frac1{2m_e^2}I A^2 g_{aee}^2a_0^2 v_z^2\,,
\end{equation}
where $v_z$ is the projection of the axion velocity $\vec v$ on the bias magnetic field $\vec B =(0,0,B)$. This projection varies from 0 to $v=10^{-3}c$ resulting in daily and annual modulations of the signal. 

Substituting the matrix element (\ref{Velement}) into Eq.~(\ref{W46}) and using the identities $a_0^2 = 2\rho_{\rm DM}/m_a^2$ and $\Gamma = \gamma m_a$, we find
\begin{equation}
    W = \frac{IA^2  \rho_{\rm DM}v_z^2}{\gamma m_a^3 m_e^2}g_{aee}^2\,.
\end{equation}
It is convenient to represent this transition rate in the parameter-independent form,
\begin{equation}
    W = W_0 g_{aee}^2 \left(\frac{\rho_{\rm DM}}{0.4\,{\rm GeV/}{\rm cm}^3} \right)
    \left( \frac{10^{-6}}{\gamma} \right)
    \left( \frac{v_z}{10^{-3}c} \right)^2,
\end{equation}
where
\begin{equation}
    W_0 = \frac{IA^2}{m_a^3 m_e^2}\frac{0.4\, {\rm GeV}}{{\rm cm}^3}\,.
\end{equation}

Consider now the experiment with $N_{\rm at}=10^{10}$ trapped atoms in the magnetic field $B$ which may be tuned for scanning for the resonance. The probability of transition within the integration time $t_{\rm int} =10$ s after $N_{\rm rep} = 10^4$ repetitions would be
\begin{equation}
    P = W_0 t_{\rm int}N_{\rm at}N_{\rm rep}g_{aee}^2\,.
\end{equation}
Assuming that this probability is of order of 1, we find the sensitivity of this experiment to the axion-electron coupling constant $g_{aee}$ in the region $1\,\mu{\rm eV}<m_a<100\,\mu{\rm eV}$:
\begin{equation}
    g_{aee}\lesssim 3.5\times 10^{-10}\,.
    \label{gaee}
\end{equation}
This constraint assumes non-detection of hyperfine atomic transitions in the proposed experiment.

The expected sensitivity (\ref{gaee}) is about 20 times lower than the existing laboratory limits on this coupling \cite{QUAX} and about 2000 times lower than the astrophysical constraints \cite{Raffelt20}. However, the sensitivity of atomic experiments may be potentially raised by increasing the number of atoms upon development of the experimental techniques. Thus, this experiment may provide independent limits on the axion-electron interaction.

Note that in our estimates we considered only the transitions between specific Zeeman sublevels $|\psi_\pm\rangle$ which are suitable for detecting the scalar field dark matter. The axion-electron interaction, however, can cause transitions to other sublevels with $\Delta m_F = \pm1$. These transitions have similar transition rates to the one considered above, but are proportional to other axion velocity components, $v_x^2$ and $v_y^2$. Therefore, detecting transitions to all these sublevels and studying their daily and annual modulations may help identify the axion dark matter and distinguish it from the scalar field.


\section{Summary}
\label{SecSummary}

If the dark matter is represented by a light scalar field with a dilaton-like interaction with the Standard Model fields, it should manifest itself in small oscillations of fundamental constants. It has been known for quite a while \cite{Arvanitaki2015,Stadnik2015} that these oscillations may cause observable variations of frequencies of atomic transitions if the scalar field mass is significantly lower than the frequencies of atomic or molecular transitions. If the mass of the scalar field appears close to frequencies of atomic transitions, the oscillating fundamental constants can serve as a source of these transitions. In this paper, we propose an experiment with cold atoms that would allow one to detect the scalar field dark matter if its mass falls within the range of hyperfine transitions in atoms.

For simplicity, we focus on alkali atoms in the $s_{1/2}$ ground state with the hyperfine splitting proportional to the magnetic dipole hyperfine constant $A$. We show that this constant may oscillate if the dark matter is represented by a light scalar field with dilaton-like interaction with the Standard Model fields. However, the oscillation of the hyperfine constant $A$ alone is not sufficient to induce hyperfine transitions in atoms because the operator of hyperfine interaction has only diagonal matrix elements in the basis $|F,m_F\rangle$. We show that such transitions are possible in the presence of external magnetic field $B$ due to the additional oscillations of the Bohr magneton (oscillation of Bohr magneton in the sourse  also leads to oscillation of the magnetic field). We calculate the corresponding resonance transition rate and show that it is proportional to the second power of the difference between the scalar field couplings involved in hyperfine and Zeeman interactions. Thus, the proposed experiment may be sensitive to a specific combination of couplings $\Lambda_{A\mu}$ given by Eq.~(\ref{LambdaAmu}). The strength of the magnetic field $B$ may be used for tuning the splitting between Zeeman sublevels and scanning for the resonance with the oscillating scalar field dark matter. 

It is noteworthy that the hyperfine transitions induced by the oscillating fundamental constant may be observed  in cold atoms, at temperatures below 1 K. It is technologically available to keep up to $N_{\rm at}=10^{10}$ of cold atoms loaded from a magneto-optical trap in a deep optical lattice for a time of measurements. We show that an experiment employing this number of cold atoms may be sensitive to $\Lambda_{A\mu}$ ranging approximately from $5\times 10^7$ GeV to $5\times 10^8$ GeV in the interval of the scalar field frequencies from 1 $\mu$eV to 100 $\mu$eV. In Fig.~\ref{FigSensitivity}, we present plots of projected sensitivity of this experiment to the couplings with photon $\Lambda_\gamma$, electron $\Lambda_e$ and proton $\Lambda_p$ assuming that either of these couplings dominates. We discuss also the sensitivity of this experiment to the Higgs field coupling $\Lambda_H$, see Fig.~\ref{FigSensitivity} (d).

The proposed experiment may provide new constraints on the coupling of the scalar field to the standard model particles if the scalar field mass appears within the interval from 1 to 100 $\mu$eV. This interval is only partly constrained for $\Lambda_\gamma$ in the range of ADMX frequencies, $2.6\, \mu\text{eV}<m_\phi < 4.2\,\mu\text{eV}$ \cite{FMST}, while other couplings do not have direct constraints. There are, however, indirect constraints on all these couplings from the equivalence principle derived in Refs.~\cite{Hees2018,Leefer2016}. Although the latter constraints are much stronger than the expected sensitivity of the proposed here experiment, it is still interesting to provide a direct and independent study of all scalar field couplings with the use of atomic hyperfine transitions. We hope these measurements will be conducted in future experiments.

In this paper, we focus on the linear coupling of the scalar field with the Standard Matter fields while quadratic couplings introduced in Refs.~\cite{Stadnik2015,Stadnik2015a,Stadnik2016,Hees2018,Stadnik2019,Kim2022} are of interest as well. Note that in this case  $\phi$ may be scalar or pseudoscalar (axion) field. It is straightforward to generalize the results of the present work to the scalar field model with quadratic couplings making use of the identity $2\cos^2\omega t = 1+ \cos 2\omega t$. This identity means that the frequency is doubled, and the constraints on the quadratic couplings may be found from the corresponding constraints on the linear ones.

We demonstrate that the proposed experiment may be sensitive to the axion-electron coupling $g_{aee}$. Although the estimated sensitivity is about 20 times lower than the existing limits from the laboratory experiments \cite{QUAX} and about 2000 times lower than the astrophysical constraints \cite{Raffelt20}, this experiment may provide independent limits on this coupling. The signal from the axion dark matter should have strong daily  modulation due change of the angle between axion momentum and magnetic field producing Zeeman splitting.

\vspace{3mm}
\textit{Acknowledgements} --- 
IBS is grateful to D.~Budker and O.~Tretiak for useful discussions. The work of VVF and IBS was supported by the Australian Research Council Grants No. DP190100974 and DP200100150. The work of CW was supported by the Cluster of Excellence `CUI: Advanced Imaging of Matter’ of the Deutsche Forschungsgemeinschaft (DFG) - EXC 2056 - project ID 390715994 and by the European Research Council (ERC) under the European Union’s Horizon 2020 research
and innovation programme under grant agreement No. 802701.


\appendix
\section{Coherence time of the scalar field dark matter}
\label{AppA}

\subsection{Transition rate under monochromatic periodic perturbation}

Let $V(t)$ be an operator of a time-dependent periodic perturbation of the form
\begin{equation} \label{Vperturb}
    V(t) = V_0 \cos \omega_\star t\,,
\end{equation}
where $V_0=V_0^\dag$ is a ``small'' time-independent operator. Let ${\cal E}_k$ be stationary energy levels of the atom, and the frequency $\omega_{k0} = \hbar^{-1}({\cal E}_k - {\cal E}_0)$ appears close to the perturbation frequency $\omega_\star$. Then, the transition rate from the ground state $|0\rangle$ to a stationary excited state $|k\rangle$ is given by Fermi's golden rule
\begin{equation}
    W_{0\to k} = \frac{\pi}2 |\langle k|V_0|0 \rangle|^2 \delta({\omega_\star - \omega_{k0}})\,.
    \label{Wmono}
\end{equation}
Note that, traditionally (see, e.g., \cite{Landau1981Quantum}), this equation has the factor $2\pi$ rather than $\pi/2$ because the time-dependent perturbation operator is usually written as $V(t) = V_0 e^{-i\omega_\star t} + V_0^\dagger e^{i\omega_\star t}$. In our case, it is convenient to use the perturbation operator in the form (\ref{Vperturb}) corresponding to the variations of fundamental constants (\ref{deltaalpha}) and (\ref{deltam}).

\subsection{Transition rate under non-monochromatic perturbation}

In reality, the time-dependent perturbations are usually non-monochromatic, but are specified by a distribution with a finite width. In this case, time-dependent perturbation is represented by the following Fourier expansion
\begin{equation}
    V(t) = V_0 \int_0^\infty d\omega \, g(\omega) \cos \omega t \,,
\end{equation}
with $g(\omega)$ the normalized spectral density, $\int_{-\infty}^{\infty} g(\omega) d\omega =1$.
In this case, the analog of Eq.~(\ref{Wmono}) is (see, e.g., \cite{QuantumOptics})
\begin{equation}
    W_{0\to k} = \frac{\pi}2 |\langle k|V_0|0 \rangle|^2 g(\omega_{k0})\,.
    \label{Wbroad}
\end{equation}
This equation reduces to (\ref{Wmono}) for $g(\omega) = \delta(\omega - \omega_\star)$.

The equation (\ref{Wbroad}) assumes that the distribution $g(\omega)$ is localized near the transition frequency $\omega_{k0}$. In particular, it may be modelled by the Maxwellian distribution,
\begin{equation}
    g(\omega) = \frac1{\sqrt{2\pi}\sigma} e^{-\frac{(\omega-\omega_{k0})^2}{2\sigma^2}}
    \label{g-distribution}
\end{equation}
centered around $\omega_{k0}$ with variance $\sigma$. In this case, 
\begin{equation}
    g(\omega_{k0}) = \frac1{\sqrt{2\pi}\sigma}\,,
\end{equation}
and Eq.~(\ref{Wbroad}) becomes
\begin{equation}
       W_{0\to k} = \frac{ |\langle k|V_0|0 \rangle|^2}{\Gamma}\,,
       \label{Wrate}
\end{equation}
where 
\begin{equation}
\label{A8}
    \Gamma = \frac{2\sqrt2\sigma}{\sqrt{\pi}}
\end{equation}
is, by definition, the width of the dark matter frequency distribution. The latter may be found from the dark matter particle velocity distribution.

\subsection{Dark matter velocity distribution}

Let $\vec v$ be the velocity of DM particle near the Earth in the Galaxy frame. In the Standard Halo Model (see, e.g., Ref.~\cite{DMdistribution}), the velocity distribution of DM particles is Maxwellian,
\begin{equation}
    f(\vec v) = \frac1{\pi^{3/2}v_0^3} e^{-v^2/v_0^2}\,,
    \label{f-Galaxy}
\end{equation}
where $v_0$ is the most probable velocity. This velocity is usually taken equal to the Galaxy rotation velocity near the Sun, $v_0 = 230$ km/s. The distribution (\ref{f-Galaxy}) is normalized
\begin{equation}
    \int f(\vec v) d^3v =1\,.
\end{equation}
Here, for simplicity, we integrate from 0 to $\infty$, although, in general, we have to integrate up to the escape velocity $v_{\rm esc} \approx 600$ km/s. Setting the upper limit of integration to $v_{\rm esc}$ would slightly change the normalization constant of the distribution function (\ref{f-Galaxy}), but in our estimates we ignore this effect.

In the laboratory frame, the distribution (\ref{f-Galaxy}) turns into
\begin{equation}
    f(\vec v, \vec v_E) = \frac1{\pi^{3/2}v_0^3} e^{-(\vec v + \vec v_E)^2/v_0^2}\,,
\end{equation} 
where $v_E$ is the velocity of the Earth in the Galaxy frame. Ignoring the rotation of the Earth around the Sun, this velocity is approximately equal to the velocity of the Sun in the Galaxy frame, $v_E\approx v_0$.

We are considering the model of cold dark matter with non-relativistic particles with mass $m$ and dispersion relation
\begin{equation}
    {\cal E}(v) = \sqrt{m^2 + p^2} \approx m+ \frac{mv^2}2\,.
\end{equation}
The average energy of these particles and the variance squared are
\begin{eqnarray}
    \bar {\cal E} &=& \int {\cal E}(v) f(\vec v,\vec v_0) d^3v = m + \frac 54 mv_0^2 
    \nonumber\\
    &\approx& m(1+7\times 10^{-7})\,,\\
    \sigma^2 &=& \overline{({\cal E}-\bar {\cal E})^2} 
    =\frac{m^2}{4} \int (v^2 - \frac 54 v_0^2)^2 f(\vec v,\vec v_0) d^3v 
    \nonumber\\
    &=& \frac{81}{64}m^2v_0^4\,.    
    \label{Ebar}
\end{eqnarray}
As a result, we find 
\begin{equation}
    \sigma = \frac98 mv_0^2 \approx 6.6\times 10^{-7}m\,.
    \label{sigma}
\end{equation}
Substituting this variance into Eq.~(\ref{A8}), we find the width of the scalar field frequency distribution 
\begin{equation}
    \Gamma \approx 10^{-6}m\,.
\end{equation}
The corresponding coherence time is 
\begin{equation}
    \tau = \Gamma^{-1} = \frac{10^{6}}{m}\,.
\end{equation}


%

\end{document}